\documentclass[twocolumn,showpacs,superscriptaddress,amsmath,amssymb,aps,prl]{revtex4-1}
\usepackage{graphicx}
\usepackage{dcolumn}
\usepackage{amsmath,bm}
\usepackage{epstopdf}
\usepackage[mathlines]{lineno}
\usepackage{color}
\usepackage[colorlinks=true,linkcolor=blue,citecolor=blue]{hyperref}

\begin{document}
\title{An exactly solvable model for a strongly spin-orbit-coupled nanowire quantum dot}
\author{Rui Li}
\email{rl.rueili@gmail.com}
\affiliation{Quantum Physics and Quantum Information Division, Beijing Computational Science Research Center, Beijing 100094, China}

\author{Lian-Ao Wu}
\affiliation{Department of Theoretical Physics and History of Science, University of the Basque Country UPV/EHU, 48008, Spain}
\affiliation{IKERBASQUE, Basque Foundation for Science, 48011 Bilbao, Spain}

\author{Xuedong Hu}
\affiliation{Department of Physics, University at Buffalo, SUNY, Buffalo, New York 14260, USA}

\author{J.~\!Q. You}
\email{jqyou@csrc.ac.cn}
\affiliation{Quantum Physics and Quantum Information Division, Beijing Computational Science Research Center, Beijing 100094, China}

\begin{abstract}
In the presence of spin-orbit coupling, quantum models for semiconductor materials are generally not exactly solvable. As a result, understanding of the strong spin-orbit coupling effects in these systems remains poor. Here we develop an analytical method to solve the one-dimensional hard-wall quantum dot problem in the presence of strong spin-orbit coupling and magnetic field, 
which allows us to obtain exact eigenenergies and eigenstates of a single electron.  With the help of the exact solution, we demonstrate some unique effects from the strong spin-orbit coupling in a semiconductor quantum dot, in particular the anisotropy of the electron g-factor and its tunability.
\end{abstract}
\date{\today}
\maketitle

{\it Introduction.}---Semiconductor materials with strong spin-orbit coupling (SOC) have attracted increasing attention because of their importance in both fundamental science and potential practical applications.  Notable examples include topological insulator phase~\cite{Hasan} in strongly spin-orbit coupled HgTe/CdTe quantum wells~\cite{Konig}, and Majorana-fermion edge states in a semiconductor nanowire~\cite{Fu,Lutchyn,Oreg} in proximity to a conventional s-wave superconductor.   Strong SOC has also been used to electrically manipulate spin qubits confined in semiconductor quantum dots (QDs)~\cite{Nowack,Nadj1,Rashba1,Golovach,Li}.  Furthermore, the Rashba SOC~\cite{bychkov} is tunable by an external electric field~\cite{Nitta}, which provides a versatile platform for exploring physical effects in the strong SOC regime. In previous studies, however, SOC is usually treated perturbatively, which is clearly not ideal if it is strong.  An exactly solvable theoretical model would no doubt help clarify the physical picture of the strong SOC limit~\cite{Echeverria,Naseri,Bulgakov,Tsitsishvili,Rashba2}.


The solution of a quantum model is determined by the corresponding Sch\"{o}dinger equation with appropriate boundary conditions.  Recently, an exact solution to the quantum Rabi model (a two-level system coupled to a harmonic oscillator) was obtained from a boundary condition based on the $\mathbb{Z}_{2}$ symmetry of the model~\cite{Braak,Xie}.  Moreover, one can map the quantum Rabi model into the problem of one electron confined in a one-dimensional (1D) harmonic QD with SOC and magnetic field.  Considering that the harmonic and hard-wall confinement potentials are two of the best known exactly solvable problems for a particle without SOC, it is natural to ask whether an exact solution for the hard-wall potential in the presence of SOC can be found, even though they appear to be quite different: one potential has boundaries at definitive positions, while the other does not.

In this Letter, we show that the problem of one electron confined in a 1D hard-wall QD with strong spin-orbit interaction in a magnetic field is exactly solvable.  We expand the Hilbert space with the eigenvectors of a harmonic oscillator, and derive a recursion relation for the expansion coefficients of the electron wave function. By imposing the hard-wall boundary condition, we derive a set of transcendental equations to obtain eigenenergies of the electron.  We then use this solution to study strong SOC effects in a semiconductor QD.  In particular, we show that the strong SOC modulates the electron g-factor in a semiconductor QD, giving rise to a periodic response to the magnetic-field direction. Furthermore, the g-factor can be tuned by adjusting either the direction of the magnetic field or the size of the QD.



{\it Nanowire QD with a Zeeman field perpendicular to the spin-orbit field.}---Consider a conduction electron that is confined in a thin semiconductor nanowire QD, and subject to both an internal SOC field and an external magnetic field. We first study the case where the magnetic field is perpendicular to the spin-orbit field and parallel to the quantum wire.  The model Hamiltonian for this case is~\cite{Gambetta}
\begin{equation}
H=\frac{p^{2}}{2m_{e}}+\alpha\sigma^{z}p+\frac{\Delta}{2}\sigma^{x}+V(x),\label{Eq_Hamiltonian}
\end{equation}
where $p=-i\hbar\partial_{x}$ is the canonical momentum, $m_{e}$ is the effective electron mass, $\alpha$ is the Rashba SOC coefficient~\cite{bychkov}, $\Delta=g_{e}\mu_{B}B$ is the Zeeman splitting, with $g_{e}$ and $\mu_{B}$ being the effective electron g-factor and the Bohr magneton, respectively. For the QD confinement potential $V(x)$, we consider a 1D square-well potential of infinite barrier height, i.e., $V(x)=0$ for $|x|<L$ and $V(x)=\infty$ for $|x|>L$, where $L$ is the size of the QD.  The nominal material for the QD is InSb, and the Hamiltonian parameters are explicitly listed in Table~\ref{Tab_Parameters}.
\begin{table}
\caption{\label{Tab_Parameters}Parameters of an InSb QD}
\begin{ruledtabular}
\begin{tabular}{ccccc}
$m_{e}/m$\footnote{$m$ is the electron mass}&$x_{\rm so}$\footnote{$x_{\rm so}=\hbar/(m_{e}\alpha)$} (nm)&$g_{e}$&$B~({\rm Tesla})$&$L $ (nm)\\
0.0136&50-200&-50.6&0.3&50
\end{tabular}
\end{ruledtabular}
\end{table}

Hamiltonian (\ref{Eq_Hamiltonian}) exhibits a $\mathbb{Z}_{2}$ symmetry, for which the combined operator $\sigma^{x}\mathcal{P}$ is a conserved quantity, where $\mathcal{P}$ is the parity operator: $[\sigma^{x}\mathcal{P},H]=0$.
%
%
Thus $\sigma^{x}\mathcal{P}$ and the Hamiltonian $H$ have common eigenfunctions $\Psi=[\Psi_{1}(x),\Psi_{2}(x)]^{\rm T}$, where $\Psi_{1}(x)$ and $\Psi_{2}(x)$ are the two spin components of a given common eigenfunction. The eigenvalue equation of the operator $\sigma^{x}\mathcal{P}$ leads to
\begin{equation}
\Psi_{1}(x)=\pm\Psi_{2}(-x),\label{Eq_relation}
\end{equation}
where $\pm$ is from $\sigma^{x}\mathcal{P}=\pm1$.  In the meantime, the Schr\"{o}dinger equation $H\Psi=E\Psi$ can be expressed in terms of $\Psi_1$ and $\Psi_2$ as
\begin{equation}
\left[\frac{p^{2}}{2m_{e}}+\alpha\,p-E\right]\Psi_{1}(x)+\frac{\Delta}{2}\Psi_{2}(x)=0.~~\label{Eq_eigenvalue_equation}
\end{equation}
In fact, the Schr\"{o}dinger equation $H\Psi=E\Psi$ in the $[\Psi_{1}(x),\Psi_{2}(x)]^{\rm T}$ space corresponds to two equations.  However, only one [e.g., Eq.~(\ref{Eq_eigenvalue_equation})] is independent, while the other can be derived by combining Eqs.~(\ref{Eq_relation}) and (\ref{Eq_eigenvalue_equation}).

\begin{figure}
\includegraphics{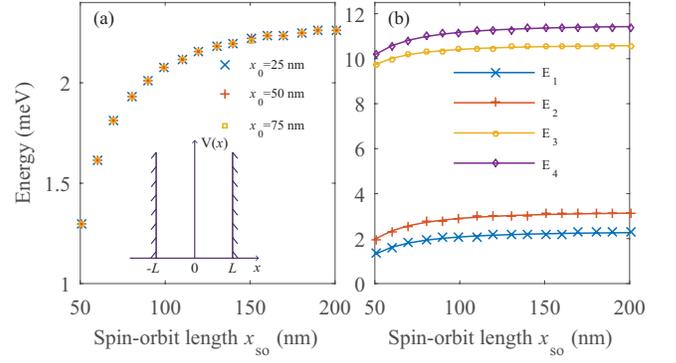}
\caption{\label{Fig_energyspectrum}Energy spectrum of the QD obtained by solving Eq.~(\ref{Eq_energy_relation}). (a) The ground-state energy of the QD calculated using harmonic-oscillator expansion basis states with different characteristic lengths: $x_{0}=25$, $50$, and $75$ nm. Inset: the hard-wall confining potential of the considered QD. (b) The lowest four energy levels of the QD.}
\end{figure}

The solution of the original single-electron problem can now be obtained by solving Eqs.~(\ref{Eq_relation}) and (\ref{Eq_eigenvalue_equation}).  We first choose a complete set of basis states, with which the two components $\Psi_{1}(x)$ and $\Psi_{2}(x)$ can be expanded.  For the QD model considered here, we find that it is essential to choose the eigenfunctions of the harmonic oscillator, $\left(\frac{p^{2}}{2m_{e}}+\frac{1}{2}m_{e}\omega^{2}_{0}x^{2}\right)\phi_{n}(x)=(n+\frac{1}{2})\hbar\omega_{0}\phi_{n}(x)$, as the basis states.  These states can be expressed as~\cite{Griffiths}
\begin{equation}
\phi_{n}(x)=1/\left(\pi\,x^{2}_{0}\right)\mathcal{H}_{n}(x/x_{0}){\rm exp}\left[-x^{2}/(2x^{2}_{0})\right] \,, \label{Eq_basis}
\end{equation}
where $\mathcal{H}_{n}(x/x_{0})$ are the normalized Hermite polynomials, and $x_{0}=\sqrt{\hbar/(m_{e}\omega_{0})}$ is the characteristic length of the harmonic oscillator.  With this basis, the matrix element of the spin-orbit interaction is non-vanishing only when $\Delta n = \pm 1$, so that the total Hamiltonian is a sparse matrix, and the iterative equations for the expansion coefficients have an automatic cut-off.  Note that $x_{0}$ (or $\omega_{0}$) is not a model parameter of Hamiltonian (\ref{Eq_Hamiltonian}), so that the energy spectrum of the QD model should not depend on $x_{0}$ (or $\omega_{0}$), c.f. Fig.~\ref{Fig_energyspectrum}.

Without loss of generality, we can expand $\Psi_{1}(x)$ as
\begin{equation}
\Psi_{1}(x)=\sum^{\infty}_{n=0}i^{n}d_{n}\phi_{n}(x),\label{Eq_varphi1}
\end{equation}
where $d_{n}$ are the expansion coefficients to be determined, and the introduction of imaginary unit $i$ is to ensure that $d_{n}$ is real. According to Eq.~(\ref{Eq_relation}), we can write $\Psi_{2}(x)$ as
\begin{equation}
\Psi_{2}(x)=\sum^{\infty}_{n=0}\pm(-i)^{n}d_{n}\phi_{n}(x),\label{Eq_varphi2}
\end{equation}
where we have used the parity property of the eigenfunction of the harmonic oscillator, i.e., $\phi_{n}(-x)=(-1)^{n}\phi_{n}(x)$. Substituting $\Psi_{1}(x)$ and $\Psi_{2}(x)$ in Eq.~(\ref{Eq_eigenvalue_equation}) with the expansions given in Eqs.~(\ref{Eq_varphi1}) and (\ref{Eq_varphi2}), we obtain the following recursion relation for the expansion coefficients~\cite{SM}:
\begin{eqnarray}
&&\sqrt{(n-1)n}d_{n-2}+\eta\sqrt{8n}d_{n-1}+\big[2n+1-4\varepsilon\pm(-1)^{n}2\xi\big]\nonumber\\
&&\times\,d_{n}+\eta\sqrt{8(n+1)}d_{n+1}=-\sqrt{(n+1)(n+2)}d_{n+2},~~~\label{Eq_iteration}
\end{eqnarray}
where $\varepsilon=E/(\hbar\omega_{0})$, $\xi=\Delta/(\hbar\omega_{0})$, and $\eta=x_{0}/x_{\rm so}$, with $x_{\rm so}=\hbar/(m_{e}\alpha)$ being the spin-orbit length.  We now let $d_{0}=1$ and introduce a new real variable $\chi=d_{1}/d_{0}$.  Now all the expansion coefficients can be determined as $d_{n}\equiv\,d_{n}(\varepsilon,\chi)$, and both $\Psi_{1,2}(x)$ are obtained in terms of these coefficients.  The energy spectrum of the Hamiltonian (\ref{Eq_Hamiltonian}) can then be straightforwardly determined by imposing the hard-wall boundary condition: $\Psi_{1,2}(\pm\,L)=0$.  In terms of the coefficients $d_n$, the boundary condition can be expressed as~\cite{SM}
\begin{eqnarray}
d_{0}\mathcal{H}_{0}-d_{2}\mathcal{H}_{2}+d_{4}\mathcal{H}_{4}-d_{6}\mathcal{H}_{6}\cdots=0,\nonumber\\
d_{1}\mathcal{H}_{1}-d_{3}\mathcal{H}_{3}+d_{5}\mathcal{H}_{5}-d_{7}\mathcal{H}_{7}\cdots=0,\label{Eq_energy_relation}
\end{eqnarray}
where $\mathcal{H}_{n}\equiv\mathcal{H}_{n}(L/x_{0})$.  With two equations for only two variables $\varepsilon$ and $\chi$, the energy spectrum can be exactly solved from Eq.~(\ref{Eq_energy_relation}).  In addition, using the property of the Hermite polynomial~\cite{Griffiths}, we can obtain the recursion relation for $\mathcal{H}_{n}$,
\begin{equation}
\sqrt{n+1}\mathcal{H}_{n+1}=\sqrt{2}(L/x_{0})\mathcal{H}_{n}-\sqrt{n}\mathcal{H}_{n-1}.\label{Eq_Hermite_relation}
\end{equation}

Equations~(\ref{Eq_iteration})-(\ref{Eq_Hermite_relation}) are our central results. The energy spectrum of the considered Hamiltonian is determined by solving two transcendental equations instead of solving a matrix eigenvalue problem.
As mentioned above, a key to our solution is that with the harmonic-oscillator basis, we are able to obtain the exact recursion relation in Eq.~(\ref{Eq_iteration}).

Figure \ref{Fig_energyspectrum} shows the energy spectrum of the problem as a function of the spin-orbit length $x_{\rm so}$, with panel (a) showing the ground energy while panel (b) showing the lowest four eigenenergies.  Recall that by choosing a complete set of harmonic-oscillator basis states, we have artificially introduced an additional parameter, the characteristic length (frequency) of the harmonic oscillator $x_{0}$ ($\omega_{0}$) [see Eq.~(\ref{Eq_basis})].  Since $x_{0}$ ($\omega_{0}$) is not a model parameter of Hamiltonian (\ref{Eq_Hamiltonian}), the energy spectrum of the system should be independent of this parameter. Indeed, our results in Fig.~\ref{Fig_energyspectrum}(a) verify this interesting point, that choosing different $x_{0}$ leads to the same eigenvalues for the considered Hamiltonian.
%

{\it Nanowire QD with a general magnetic field.}---We now turn to the general case where the magnetic-field direction is arbitrary in the $x$-$z$ plane. The model Hamiltonian takes the form
\begin{equation}
H=\frac{p^{2}}{2m_{e}}+\alpha\sigma^{z}p+\frac{\Delta}{2}\left(\sigma^{x}\cos\theta+\sigma^{z}\sin\theta\right)+V(x),\label{Eq_generalizedHamiltonian}
\end{equation}
where $\theta$ is the azimuthal angle of the in-plane magnetic field. In contrast to Hamiltonian (\ref{Eq_Hamiltonian}) for the special case of a perpendicular field, the Hamiltonian here no longer exhibits the $\mathbb{Z}_{2}$ symmetry.  However, as we show below, this model is still exactly solvable.

The two components $\Psi_{1}(x)$ and $\Psi_{2}(x)$ of the wave function $\Psi$ satisfy the Schr\"{o}dinger equations
\begin{eqnarray}
\left(\frac{p^{2}}{2m_{e}}+\alpha\,p+\frac{\Delta}{2}\sin\theta-E\right)\Psi_{1}+\frac{\Delta}{2}\Psi_{2}\cos\theta&=&0,\nonumber\\
\left(\frac{p^{2}}{2m_{e}}-\alpha\,p-\frac{\Delta}{2}\sin\theta-E\right)\Psi_{2}+\frac{\Delta}{2}\Psi_{1}\cos\theta&=&0.~~~
\label{Eq_eigenvalue_equation_ge}
\end{eqnarray}
We again choose the eigenfunctions of a 1D harmonic oscillator as the basis states.  In the absence of $\mathbb{Z}_{2}$ symmetry, we have to introduce two independent series of expansion coefficients for $\Psi_{1}(x)$ and $\Psi_{2}(x)$,
\begin{equation}
\Psi_{1}(x)=\sum^{\infty}_{n=0}i^{n}e_{n}\phi_{n}(x),~~\Psi_{2}(x)=\sum^{\infty}_{n=0}i^{n}f_{n}\phi_{n}(x),
\end{equation}
where $e_{n}$ and $f_{n}$ are the coefficients to be determined. Substituting the two components $\Psi_{i}$ ($i=1,2$) in Eq.~(\ref{Eq_eigenvalue_equation_ge}), we obtain two sets of recursion relations
\begin{widetext}
\begin{eqnarray}
&&\sqrt{n(n-1)}e_{n-2}+\sqrt{8n}\eta\,e_{n-1}+(\Xi_{n,\varepsilon}
+2\xi\sin\theta)e_{n}+(-1)^{n}2\xi\cos\theta\,f_{n}+\sqrt{8(n+1)}\eta\,e_{n+1}
=-\sqrt{(n+1)(n+2)}e_{n+2},\nonumber\\
&&\sqrt{n(n-1)}f_{n-2}+\sqrt{8n}\eta\,f_{n-1}+(\Xi_{n,\varepsilon}+2\xi\sin\theta)f_{n}
+(-1)^{n}2\xi\cos\theta\,e_{n}+\sqrt{8(n+1)}\eta\,f_{n+1}=-\sqrt{(n+1)(n+2)}f_{n+2},
\nonumber\\\label{Eq_iteration_general}
\end{eqnarray}
\end{widetext}
where $\Xi_{n,\varepsilon}=2n+1-4\varepsilon$.  While these recursion relations are much more complex than the one for the special case of Hamiltonian (1) with $\mathbb{Z}_{2}$ symmetry, we can follow the same procedure to solve them.  Specifically, we let $e_{0}=1$, and introduce three new real variables, $e_{1}/e_{0}=\chi_{1}$, $f_{0}/e_{0}=\chi_{2}$, and $f_{1}/e_{0}=\chi_{3}$.  Now all the expansion coefficients can be determined as functions of these three parameters together with the rescaled energy $\varepsilon = E/\hbar\omega_0$: $e_{n} \equiv \,e_{n}(\varepsilon,\chi_{1}, \chi_{2}, \chi_{3})$ and $f_{n} \equiv \,f_{n}(\varepsilon, \chi_{1}, \chi_{2}, \chi_{3})$.  Compared to the simpler Hamiltonian (\ref{Eq_Hamiltonian}) with $\mathbb{Z}_{2}$ symmetry, we now need to introduce three new real variables $\chi_1$, $\chi_2$, and $\chi_3$, instead of one new real variable $\chi$.

The energy spectrum of the system can again be determined by the hard-wall boundary condition $\Psi_{1,2}(\pm\,L)=0$, which gives rise to
\begin{eqnarray}
e_{0}\mathcal{H}_{0}-e_{2}\mathcal{H}_{2}+e_{4}\mathcal{H}_{4}-e_{6}\mathcal{H}_{6}\cdots=0,\nonumber\\
e_{1}\mathcal{H}_{1}-e_{3}\mathcal{H}_{3}+e_{5}\mathcal{H}_{5}-e_{7}\mathcal{H}_{7}\cdots=0,\nonumber\\
f_{0}\mathcal{H}_{0}-f_{2}\mathcal{H}_{2}+f_{4}\mathcal{H}_{4}-f_{6}\mathcal{H}_{6}\cdots=0,\nonumber\\
f_{1}\mathcal{H}_{1}-f_{3}\mathcal{H}_{3}+f_{5}\mathcal{H}_{5}-f_{7}\mathcal{H}_{7}\cdots=0.\label{Eq_determine_ES}
\end{eqnarray}
These four equations are exact, and we have only four unknowns, $\varepsilon$, $\chi_{1}$, $\chi_{2}$, and $\chi_{3}$.  Thus the four variables can be exactly determined, and the energy spectrum of the system obtained.

\begin{figure}
\includegraphics{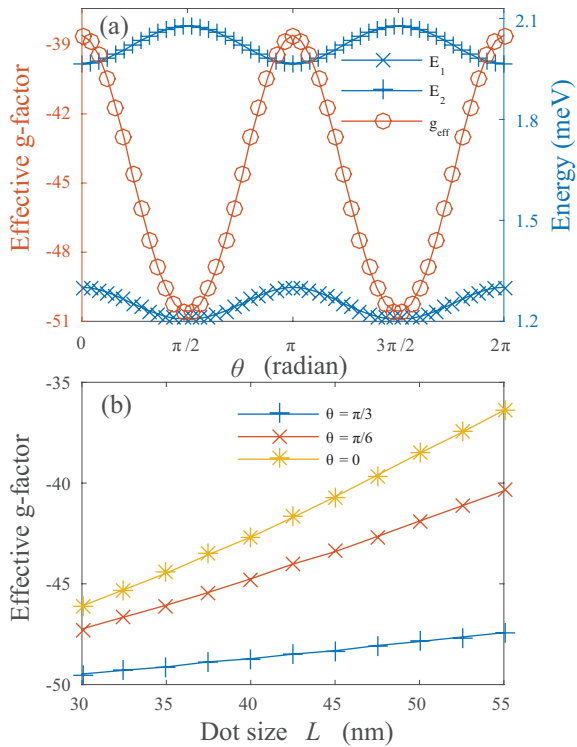}
\caption{\label{Fig_gfactor}Effective electron g-factor of the QD Hamiltonian (\ref{Eq_generalizedHamiltonian}). Here the applied magnetic field is $0.3$ T and the spin-orbit length is chosen as $x_{\rm so}=50$ nm. (a) The g-factor as a function of the applied magnetic-field direction. (b) The g-factor as a function of the QD size $L$.}
\end{figure}

The general solutions derived here converge to the special case solution when we take $\theta=0$.  Specifically, the two sets of recursion relations (\ref{Eq_iteration_general}) are reduced to just one, in the form of recursion relation (\ref{Eq_iteration}), and the boundary condition (\ref{Eq_determine_ES}) is reduced to Eq.~(\ref{Eq_energy_relation}).  Clearly, while $\mathbb{Z}_{2}$ symmetry simplifies the calculations, it is not an essential ingredient in our solution here.  Instead, it is the choice of the harmonic oscillator eigenstates as expansion basis that is crucial to our exact solutions of the model Hamiltonians (\ref{Eq_Hamiltonian}) and (\ref{Eq_generalizedHamiltonian}). This is different from the quantum Rabi model, where the $\mathbb{Z}_{2}$ symmetry plays an essential role to exactly solve the model~\cite{Braak,Xie}.

With the exact solution to the general Hamiltonian we can study a QD with strong SOC.  One of the key properties of a confined electron is its g-factor, which is strongly affected by SOC strength, and possibly by the QD confinement.  With our nominal example of InSb, its bulk g-factor is isotropic and takes the value $g_{e}=-50.6$.  Figure~\ref{Fig_gfactor} shows our calculated g-factor, which can be defined as $g_{\rm eff}=(E_{2}-E_{1})/(\mu_{B}B)$ of a single confined electron with Hamiltonian (\ref{Eq_generalizedHamiltonian}).  This relationship should be valid as long as Zeeman splitting is much smaller than the orbital excitation energy.  Figure~\ref{Fig_gfactor}(a) shows the effective g-factor as a function of the direction of the applied magnetic field, with a clear anisotropy as the field rotates~\cite{Nilsson,Schroer,Nadj2}. It is a periodic function of $\theta$ with a period of $\pi$.  In particular, for $\theta=\pi/2$ or $3\pi/2$, the operator $\sigma^{z}$ is conserved by Hamiltonian (\ref{Eq_generalizedHamiltonian}), so that the SOC effect is straightforward to obtain, and the electron g-factor is equal to the bulk value $g_{e}=-50.6$. For $\theta=0$ or $\pi$, the magnetic field is perpendicular to the spin-orbit field, so that $\sigma^{z}$ is no long conserved, resulting in the largest modulation to the electron g-factor, with $g_{\rm eff}\approx-38.6$. Figure~\ref{Fig_gfactor}(b) shows the g-factor dependence on the QD size $L$.  The SOC strength in the QD is characterized by the ratio $L/x_{\rm so}$.  When $L$ increases, reducing the orbital energy splittings, the spin-orbit induced level mixing becomes stronger, which is equivalent to an increase in the relative SOC strength.  Consequently, increasing $L$ (but still within the range when the orbital excitation energy is larger than the Zeeman splitting) enhances the modulation to the electron g-factor.  In short, Fig.~\ref{Fig_gfactor} shows that we can manipulate the electron g-factor by varying the applied magnetic-field direction and the nanowire QD size.


{\it Summary.}---In this study we have obtained the exact energy spectrum of a 1D hard-wall QD in the presence of both a strong SOC and an applied magnetic field.  The key ingredient of our solution is the selection of the bare harmonic-oscillator eigenfunctions as the basis states for the single-electron Hilbert space, which allows us to express the system Hamiltonian as a sparse matrix and obtain a recursive relation for the wave function that can be solved.  With the help of this solution, we are able to study strong SOC effects in in a 1D semiconductor QD beyond the perturbation limit.  In particular, we demonstrate a strong anisotropy in the electron g-factor, and the tuning of the electron g-factor via external parameters, i.e., the QD size and the applied magnetic-field direction.

R.L. and J.Q.Y. are supported by National Natural Science Foundation of China Grant Nos.~91421102 and 11404020, National Basic Research Program of China Grant No.~2014CB921401, NSAF Grant No.~U1330201, and Postdoctoral Science Foundation of China Grant No.~2014M560039. L.W. acknowledges grant support from the Basque Government (Grant No. IT472-10) and the Spanish MICINN (Grant No. FIS2012-36673-C03-03).  X.H. acknowledges financial support by US ARO (W911NF0910393) and NSF PIF (PHY-1104672).

\end{document}